\documentclass[conference]{IEEEtran}
\IEEEoverridecommandlockouts
\usepackage{cite}
\usepackage{amsmath,amssymb,amsfonts}
\usepackage{algorithmic}
\usepackage{graphicx}
\usepackage{textcomp}
\usepackage{xcolor}
\usepackage{subfigure}
\usepackage{multirow}
\usepackage{booktabs}
\usepackage{array}
\usepackage{color}
\usepackage{multirow}
\usepackage{makecell}
\def\BibTeX{{\rm B\kern-.05em{\sc i\kern-.025em b}\kern-.08em
    T\kern-.1667em\lower.7ex\hbox{E}\kern-.125emX}}

\usepackage{fancyhdr}
\begin{document}

\title{Non-Line-of-Sight imaging using raster scanning at NIR wavelength\\
{\footnotesize \textsuperscript{}}
\thanks{}
}

\author{
\IEEEauthorblockN{Mohammad Roueinfar}
\IEEEauthorblockA{\textit{School of Electrical Engineering } \\
\textit{Iran University of Science and Technology }\\
Tehran, Iran \\
m\_roueinfar@alumni.iust.ac.ir}

\and
\IEEEauthorblockN{Mahdi Salmanian}
\IEEEauthorblockA{\textit{School of Physics} \\
\textit{Shahid Beheshti University}\\
Tehran, Iran  \\
mahdiopo@gmail.com}
}

\maketitle
\thispagestyle{fancy}

\begin{abstract}
Non-line-of-sight (NLOS) imaging is an emerging technique with transformative potential, enabling the visualization of hidden objects through indirect light reflection. This paper presents a NLOS imaging method operating in the near-infrared (NIR) wavelengths, specifically employing a raster scanning technique with a pan-tilt device. The NIR laser, operating at a wavelength of 808 nm and an output power of 500 mW, illuminates a hidden target occluded by an obstacle. The imaging process involves three bounces: the laser beam first strikes a relay wall, then reflects off the hidden target, returns to the relay wall, and subsequently reaches the NIR camera. This study systematically evaluates the effectiveness of the proposed method across three distinct targets, demonstrating the capability to recover high-quality images from non-line-of-sight scenarios. The obtained images of the hidden targets are compared with their ground truth images, and the error in the obtained images is assessed based on the criteria of Mean Squared Error (MSE) and Root Mean Square Error (RMSE).
\end{abstract}

\section{Introduction}
Non-line-of-sight (NLOS) imaging is a rapidly growing research area over the past decade, driven by significant advancements in optical sensing technologies and computational algorithms. NLOS imaging enables the visualization of objects beyond the direct line of sight, known as hidden objects. Unlike Line-of-Sight (LOS) imaging, where reflected light directly enters the sensor, NLOS imaging relies on analyzing diffuse reflections from reflective surfaces to reconstruct hidden objects \cite{b1}. NLOS imaging has a variety of applications, including:
\begin{itemize}
\item Medical Imaging: Beneficial in endoscopic procedures for visualizing internal organs\cite{b2}.
\item Security and Defense: Essential in urban warfare for detecting hidden threats \cite{b3}. 
\item Autonomous Driving: Enhances safety by imaging blind spots and areas outside the driver's view \cite{b4}.
\item Accident Prevention: Assists drivers in avoiding collisions, particularly in low-visibility scenarios \cite{b4}.
\item Parking Assistance: Guides drivers in environments with limited visibility \cite{b5}.
\item Firefighting: Helps in locating people and hazards in smoke-filled environments \cite{b6}.
\item Search and Rescue: Assists in locating individuals in disaster-stricken areas where direct line of sight is obstructed \cite{b7}.
\item Industrial Inspection: Used for inspecting hard-to-reach areas in industrial settings, such as inside machinery or pipelines \cite{b8}.
\end{itemize}
According to \cite{b1}, the classification of NLOS imaging methods can be organized into several key categories, each with distinct characteristics and applications: Non-line-of-sight (NLOS) imaging methods can be classified into several categories based on their operational principles and reconstruction techniques. Active NLOS imaging methods utilize controllable light sources, such as picosecond pulsed lasers, to indirectly illuminate hidden scenes, enabling high-resolution 3D reconstruction through sophisticated photon detection. In contrast, passive NLOS imaging relies on ambient light or light emitted by hidden objects, often resulting in less detailed reconstructions.\\
The reconstruction algorithms can be further divided into traditional physics-based methods, which focus on the time-of-flight information of photons, and deep learning-based methods which leverage advanced algorithms for improved performance. Additionally, end-to-end deep learning models streamline the reconstruction process by directly mapping measurement data to 3D volumes, enhancing efficiency. This classification framework underscores the diverse approaches within NLOS imaging technology, which holds significant promise for autonomous driving, medical imaging, and defense applications. A number of the most important studies that have used the methods mentioned above are summarized in Table 1.
\begin{table}[h]
\centering
\caption{Literature Review Summary}
\begin{tabular}{|m{0.5cm}|m{1.5cm}|m{2.5cm}|m{2.5cm}|}
\hline
\textbf{Study} & \textbf{Imaging Setup} & \textbf{Key Contributions} & \textbf{Methods \& Algorithms} \\
\hline
\cite{b9} & Confocal and dual telescope setup & Imaging at 1.43 km & Advanced hardware, direct imaging \\
\hline
\cite{b10} & Phasor field & High-quality complex scene reconstruction & Phasor field transformation \\
\hline
\cite{b11} & NLOS setup & 3D deconvolution problem & 3D deconvolution \\
\hline
\cite{b12} & Various NLOS setups & Automatic feature extraction & Deep learning (end-to-end) \\
\hline
\cite{b13} & Various NLOS setups & Physics-based reconstruction & Deep learning (physics-based) \\
\hline
\cite{b14} & Two-bounce behind obstacles & New NLOS scene types & Two-bounce method \\
\hline
\cite{b15} & Keyhole setup & Innovative techniques & Keyhole imaging \\
\hline
\cite{b16} & LWIR wavelength setup & Far-infrared imaging, latent heat source & Single jump problem, strong reflectance \\
\hline
\cite{b17} & Far-infrared wavelength setup & Polarization and deep learning & Polarization-based deep learning \\
\hline
\cite{b18} & THz NLOS imaging & 3D spatial location & passive THz plenoptic measurements \\
\hline
\end{tabular}
\end{table}
\section{Proposed NLOS Imaging method}
\begin{figure}[h]
\centerline{\includegraphics[angle=0,width=0.45\textwidth]{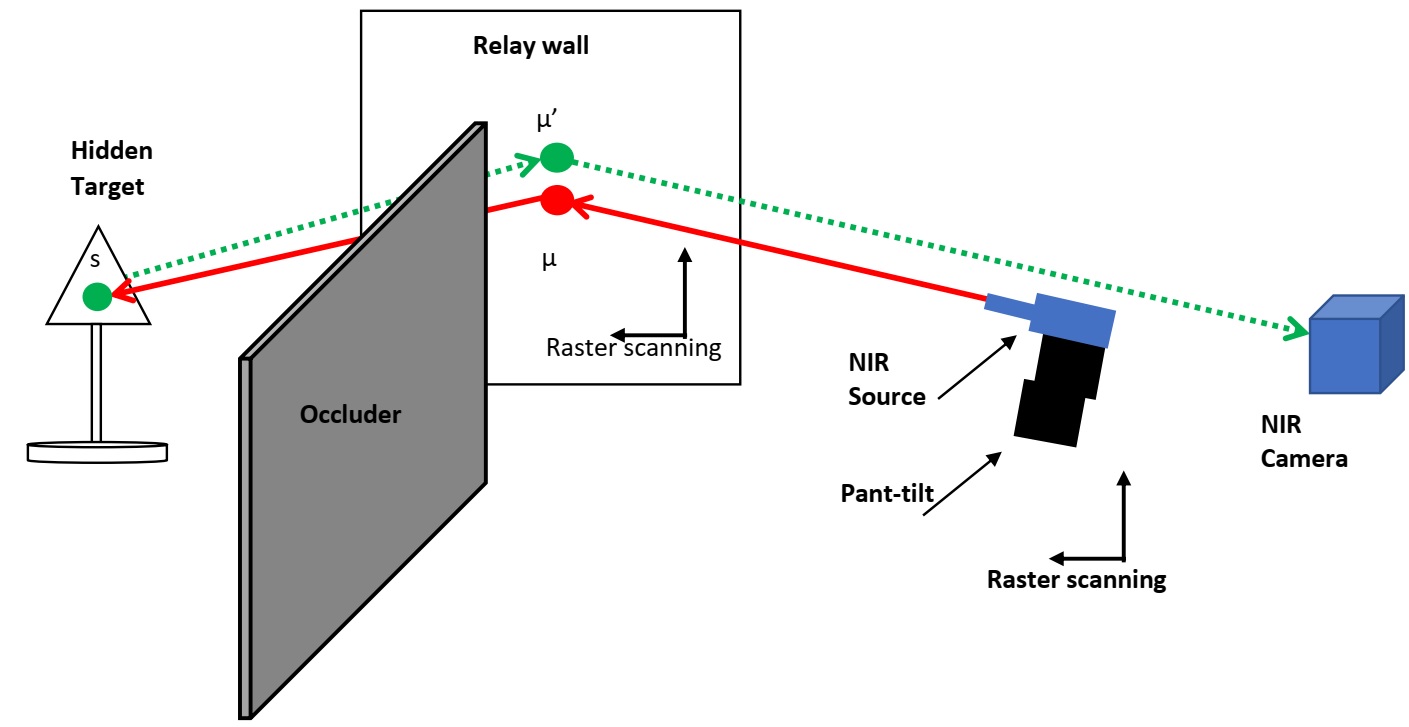}}
\caption{NLOS imaging setup.}
\label{fig1}
\end{figure}

According to \cite{b1}, the proposed Non-Line-of-Sight (NLOS) imaging method is based on the optical transmission process. The measurement data $\tau$ is defined as a function of the hidden target. As shown in ``Fig.~\ref{fig1}'', the proposed NLOS imaging system consists of an NIR laser as the emitter and an NIR camera as the sensor. The target is hidden from direct view of both the emitter and the sensor due to the presence of an obstacle or occluder between the target and the other components of the imaging system.
The occluder blocks the direct line-of-sight (LOS) path of the NIR light, resulting in non-line-of-sight (NLOS) imaging and the use of a relay wall. This imaging process involves three bounces: first, the emitted NIR laser irradiates the illumination point $\mu$  on the relay wall; second, the NIR light reflects off the hidden target surface; and third, the reflected light from the relay wall is collected at the detection point $\mu^{'}$.
For a given irradiation point $\mu$, the collected signal $\tau(\mu,\mu^{'},t)$ at time $t$ can be expressed as follows \cite{b1}:
\begin{align}
\label{eq:e1}
\tau(\mu,\mu^{'},t)= & \int \rho(s) F(\mu \rightarrow s \rightarrow \mu^{'}) \cdot \nonumber \\
& R(\mu \rightarrow  s \rightarrow \mu^{'}) B(\mu \rightarrow  s \rightarrow \mu^{'}) ds  
\end{align}
where $s$ represents a point on the hidden object, $\rho$ denotes the albedo of point $s$, $F$ refers to the optical transmission process from the irradiation point $\mu$ to the detection point $\mu^{'}$ via the hidden object surface, $R$ stands for the amplitude attenuation, and $B$ represents the bidirectional reflection distribution function (BRDF)\cite{b1}.\\
As shown in ``Fig.~\ref{fig1}'', using a pan-tilt mechanism, the hidden target surface is scanned in a two-dimensional raster pattern. The procedure involves mounting the NIR laser on the pan-tilt system, which performs the raster scan. At each scanning point, the NIR laser beam is emitted towards the relay wall, reflects off the wall, reaches the hidden target, reflects again, and then returns to the relay wall before finally entering the NIR camera. The camera records and stores the return signal $\mu^{'}$  from each scanned point. This return signal $\tau(\mu,\mu^{'},t)$ is indexed by the $n$-th scan point in the horizontal direction and the $m$-th scan point in the vertical direction. Collectively, the intensity of each return signal forms a pixel, and the sum of these intensities across different points creates an image. The final image is obtained using the following equation:
\begin{align}
\label{eq:e2}
I(n,m)=\sum_{n=1}^{N}  \sum_{m=1}^{M} \tau(n,m)
\end{align}
where $N$ is the number of horizontal scan points and $M$ is the number of vertical scan points.
Thus, the final image of the hidden object can be obtained through post-processing using suitable thresholding and other necessary techniques. The number of points that illuminate the target during raster scanning is limited. However, increasing the number of these points enhances the completeness and clarity of the image of the hidden object.
\section{EXPRIMENTAL NLOS IMAGING AT NEAR-INFRARED WAVELENGTHS}	
\begin{figure}[htbp]
\centerline{\includegraphics[angle=0,width=0.32\textwidth]{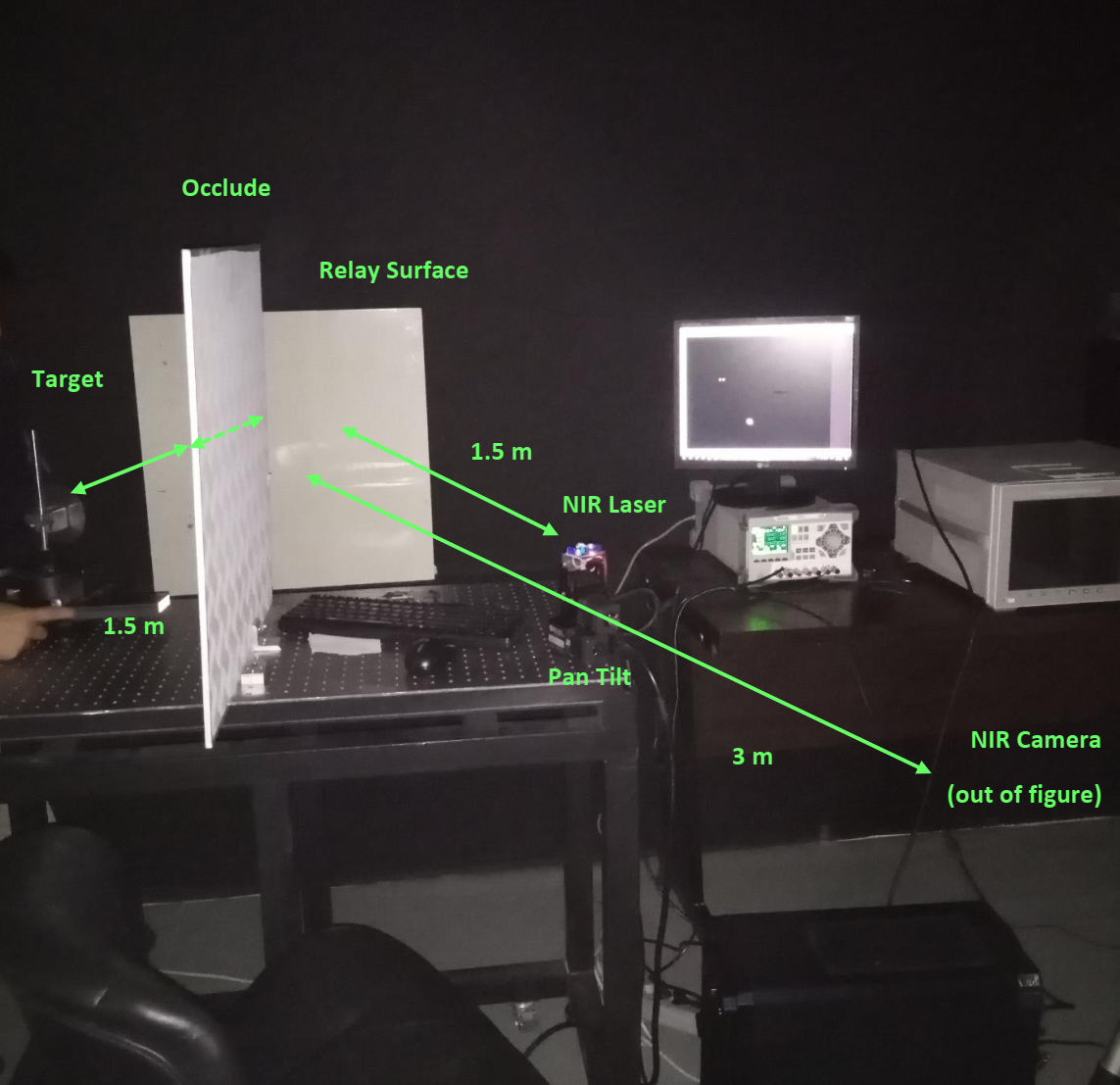}}
\caption{Image of the actual NLOS setup at NIR wavelength.}
\label{fig2}
\end{figure}

Based on the proposed NLOS imaging method from the previous section, a laboratory setup operating at the NIR wavelength has been established, as shown in ``Fig.~\ref{fig2}'' and in accordance with ``Fig.~\ref{fig1}'', the NIR laser source and NIR camera are positioned on one side of an occluder, while the hidden target is situated on the opposite side, effectively out of the direct line of sight of both the source and the camera. Utilizing a three-bounce NLOS imaging technique, the imaging of three different hidden targets is achieved with the aid of a reflective wall.
The technical specifications of the NLOS imaging components are provided in Table 2.
\begin{table}[h]
\centering
\caption{Technical Specifications of the NLOS Imaging System}
\begin{tabular}{|m{4cm}|m{3.5cm}|}
\hline
\textbf{Parameter} & \textbf{Values}  \\
\hline
Imaging method & NLOS (non-line-of-sight) \\
\hline
Band & NIR (Near-Inferared) \\
\hline
Wavelength of NIR Laser & 808 nm \\
\hline
Power of NIR Laser & 500mW \\ 
\hline
NIR Camera model & DCC1240-C \\
\hline
Resolution of NIR Camera & 1.3 Megapixels (1280 x 1024) \\
\hline
Distance from NIR Laser to Relay Wall & 1.5 m \\
\hline
Distance from Hidden Target to Relay Wall & 1.5 m \\
\hline
Distance from the NIR Camera to Relay Wall & 3 m \\
\hline
Pan-tilt Model & PTU-46-70 \\
\hline
Resolution of Pan-Tilt System & 0.012857 degrees \\
\hline
Speed of Pan-Tilt System& 300 degrees/sec \\
\hline
Scanning Type & Raster Scan \\
\hline
Scanning Grid Dimensions & 16 $\times$ 16 points\\
\hline
Relay wall Material & Ceramic \\
\hline
Hidden target material & Steel \\
\hline
Test environment & Room \\
\hline
\end{tabular}
\end{table}
As shown in ``Fig.~\ref{fig4}'', the hidden target is scanned indirectly with NIR radiation in two dimensions. The NIR return beam, reflected from the hidden target and then the reflecting wall, is recorded and stored by the NIR camera. By combining the returned signals and applying appropriate interpolation and necessary post-processing, the image of the hidden target is obtained.
\subsection{NIR Laser and Pan-Tilt Configuration}
A near-infrared (NIR) laser with a wavelength of 808 nm and a power output of 500 mW is mounted as NIR source on a pan-tilt mechanism. The ``Fig.~\ref{fig31}'' illustrates this laser in conjunction with the pan-tilt system. Specifically, ``Fig.~\ref{fig31}'' depicts the NIR laser, alongside the pan-tilt camera, which is utilized based on the available facilities. This setup is accessible to the operator of an NIR camera, specifically the DCC1240-C model.  
\begin{figure}[ht]
\centering
\begin{subfigure}[]
    \centering
    \includegraphics[angle=0,width=0.15\textwidth]{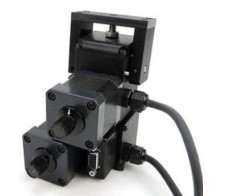}
    \label{fig31}
\end{subfigure}
\begin{subfigure}[]
    \centering
    \includegraphics[angle=0,width=0.10\textwidth]{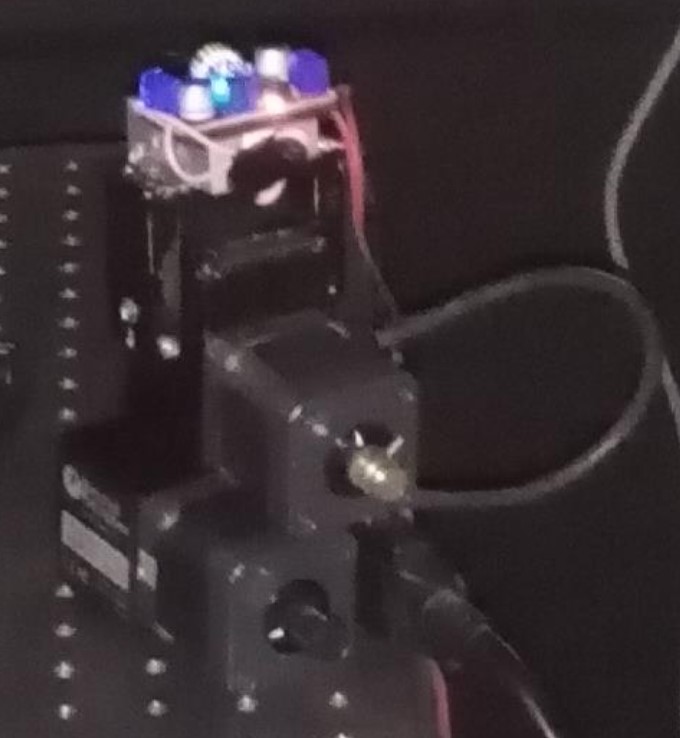}
    \label{fig32}
\end{subfigure}
  \caption{(a) Pan-Tilt without NIR source. (b)Pan-Tilt with NIR source.}
\label{fig31}
\end{figure}
We mount the NIR laser on the pan-tilt and scan a number of points in a matrix pattern as shown in the ``Fig.~\ref{fig31}''. The  ``Fig.~\ref{fig4}'' shows the path of the NIR laser beam on the occluder at each scanning point.
\begin{figure}[htbp]
\centerline{\includegraphics[angle=0,width=0.15\textwidth]{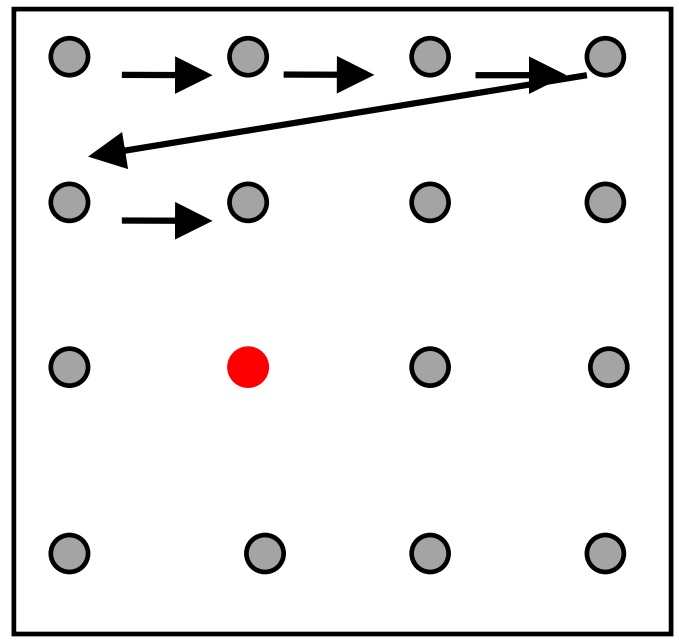}}
\caption{Scanning matrix pattern across a 4x4 grid.}
\label{fig4}
\end{figure}
By systematically scanning these points, the laser beam emits the NIR radiation onto the occluder. Upon reflection from the occluder, the reflected beam is directed towards the NIR camera. The NIR camera utilized in this laboratory setup is depicted in ``Fig.~\ref{fig12}''.
\begin{figure}[htbp]
\centerline{\includegraphics[angle=0,width=0.15\textwidth]{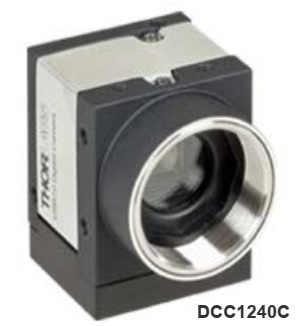}}
\caption{NIR camera.}
\label{fig12}
\end{figure}
The results obtained and the images captured by the NIR camera from raster scanning at each scan point are presented in `Fig.~\ref{fig44}''. As shown in `Fig.~\ref{fig44}'', each image depicts the transient reflection of the NIR beam as recorded and stored by the camera. In other words, each image in `Fig.~\ref{fig44}'' captures the reflection of the NIR beam at a specific moment in time. The solid dots represent a red reference laser, indicating the distance of the reflected beam from the reference point. The light dots represent the reflection of the NIR beam, which is scanned across the target surface via raster scanning and recorded by the NIR camera at each instance. 

Additionally, Fig.~\ref{fig55}'' shows the ground truth images of the hidden targets. Fig.~\ref{fig66}'' presents the corresponding images reconstructed using the NLOS imaging technique with the NIR camera. This sequence of figures effectively demonstrates the capability of the proposed NLOS imaging method to visualize hidden objects through indirect light reflection.
\begin{figure}
\centerline{\includegraphics[angle=0,width=0.28\textwidth]{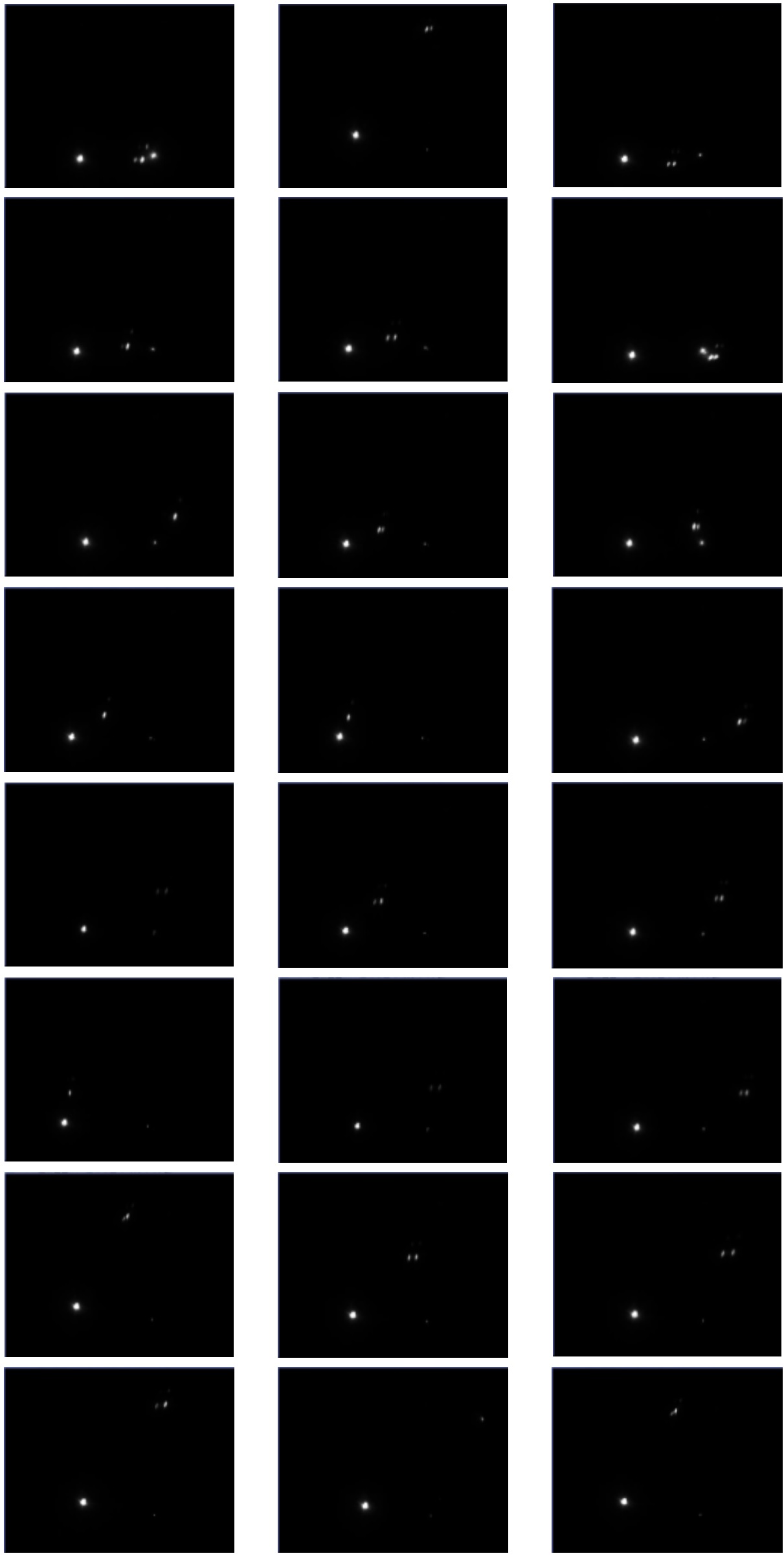}}
\caption{Each image captures the transient reflection of the NIR beam as recorded by the camera. The solid dots represent a red reference laser, indicating the distance from the reference point. The faint dots indicate the NIR beam's reflection, raster-scanned over the target surface.}
\label{fig44}
\end{figure}

\begin{figure}
\centering
\begin{subfigure}[]
    \centering
    \includegraphics[angle=0,width=0.10\textwidth]{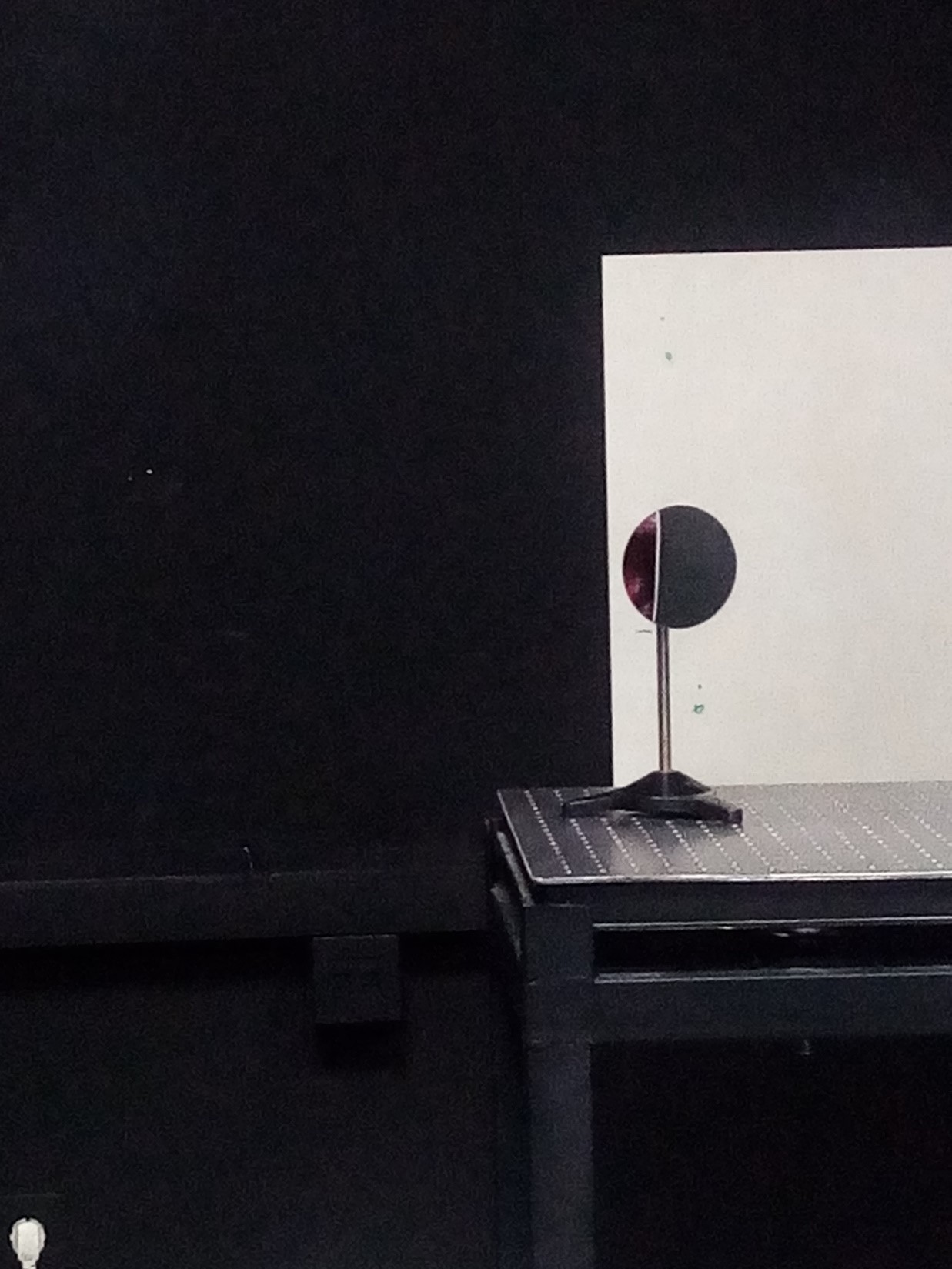}
    \label{fig5}
\end{subfigure}
\begin{subfigure}[]
    \centering
    \includegraphics[angle=0,width=0.10\textwidth]{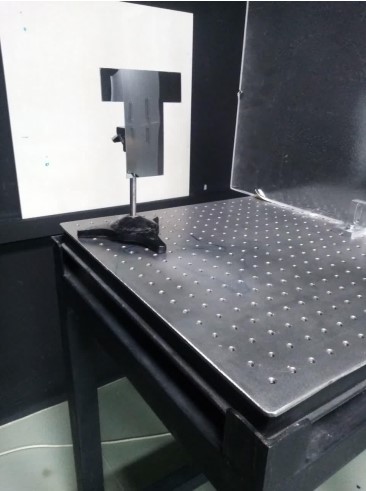}
    \label{fig7}
\end{subfigure}
\begin{subfigure}[]
    \centering
    \includegraphics[angle=0,width=0.10\textwidth]{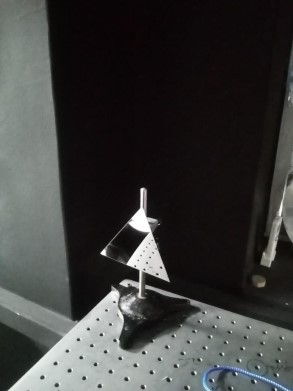}
    \label{fig9}
\end{subfigure}
  \caption{(a) A T-shaped target. (b) A circular target. (c) A triangular target.}
\label{fig55}
\end{figure}

\begin{figure}
\centering
\begin{subfigure}[]
    \centering
    \includegraphics[angle=0,width=0.3\textwidth]{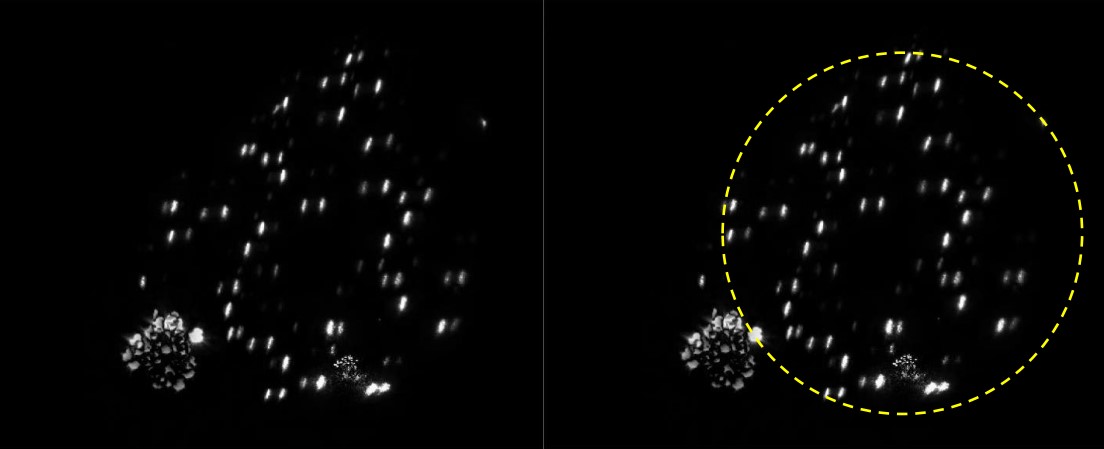}
    \label{fig5}
\end{subfigure}
\vfill
\begin{subfigure}[]
    \centering
    \includegraphics[angle=0,width=0.3\textwidth]{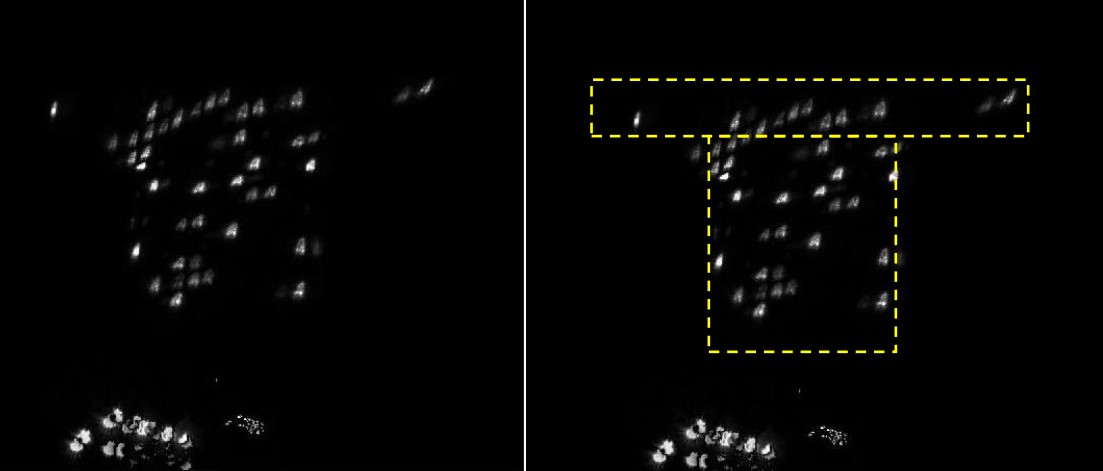}
    \label{fig7}
\end{subfigure}
\vfill
\begin{subfigure}[]
    \centering
    \includegraphics[angle=0,width=0.3\textwidth]{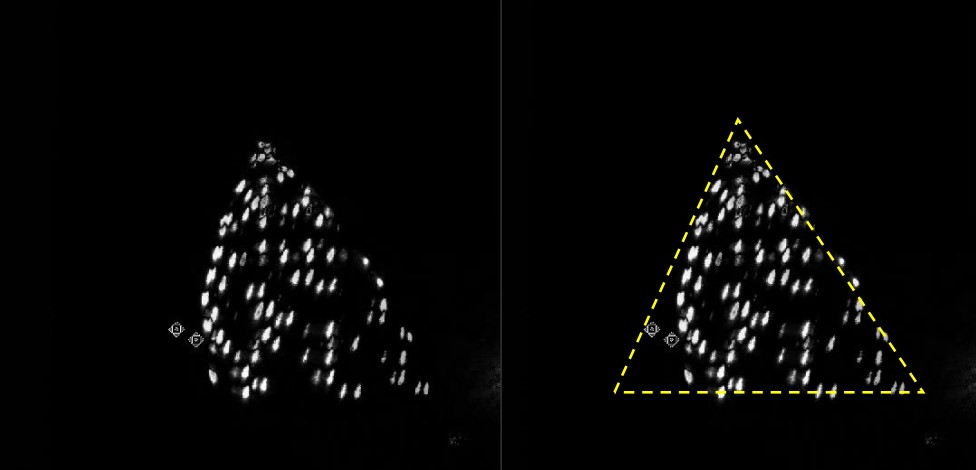}
    \label{fig9}
\end{subfigure}
  \caption{(a) The recovered image for a T-shaped target. (b) The recovered image for a circular target. (c) The recovered image for a triangular target.}
\label{fig66}
\end{figure}
The obtained image is represented as  $\boldsymbol{Z}^{'} \in \mathbb{C} ^{N \times M}$, and the ground truth image is represented as $\boldsymbol{Z} \in \mathbb{C} ^{N \times M}$. Both image matrices $\boldsymbol{Z}$ and $\boldsymbol{Z}^{'}$ are converted to vectors $\boldsymbol{z} \in \mathbb{C} ^{1 \times P}$ and $\boldsymbol{z}^{'} \in \mathbb{C} ^{1 \times P}$, respectively, where $P=N \times M$ . To compare the original image and the obtained image, two criteria, Mean Squared Error (MSE) and Root Mean Square Error (RMSE), are used, defined as follows:
\begin{equation}
\label{eq:e3}
\text{MSE} = \frac{1}{P} \sum_{i=1}^{P} (z_i - \hat{z}_i)^2
\end{equation}

\begin{equation}
\label{eq:e4}
\text{RMSE} = \sqrt{\frac{1}{P} \sum_{i=1}^{P} (z_i - \hat{z}_i)^2}
\end{equation}
In the above equations, $z_i$ represents the pixels of the original image, $\hat{z}_i$ represents the pixels of the obtained image (corresponding to the original image), and $P$ is the total number of pixels in the image. The measured values for the two criteria, MSE and RMSE, for the obtained images of the three hidden targets are presented in the table below:
\begin{table}[h]
\centering
\caption{Comparison of MSE and RMSE for different scan resolutions.}
\begin{tabular}{|m{1cm}|m{0.75cm}|m{0.75cm}|m{0.75cm}|m{0.75cm}|m{0.75cm}|m{0.75cm}|}
\hline
\textbf{Image} & \multicolumn{2}{c|}{\textbf{Scan 4x4}} & \multicolumn{2}{c|}{\textbf{Scan 8x8}} & \multicolumn{2}{c|}{\textbf{Scan 16x16}} \\ \hline
 & \textbf{MSE} & \textbf{RMSE} & \textbf{MSE} & \textbf{RMSE} & \textbf{MSE} & \textbf{RMSE} \\ \hline
\textbf{Hidden Target (a)}  & 0.4654 & 0.6822 & 0.2935 & 0.5417 & 0.1601 & 0.4001\\ \hline
\textbf{Hidden Target (b)}  & 0.5074 & 0.7132 & 0.3578 & 0.5981 & 0.1891 & 0.4348 \\ \hline
\textbf{Hidden Target (c)}  & 0.3956 & 0.6289 & 0.3069 & 0.5539 & 0.1431 & 0.3782\\ \hline
\end{tabular}
\label{table}
\end{table}

As seen in the above figures, the MSE for the three targets, although not negligible, is acceptable given the limited number of scan points in the raster scan. Naturally, increasing the number of scan points will result in a better MSE. However, it is important to note that increasing the number of scan points will also increase the time required for the raster scan and the image extraction time from the scan points, which is not desirable. Therefore, a tradeoff between the number of scan points and the MSE value should be considered.
\section{CONCLUSION}
In this paper, we introduced a non-line-of-sight (NLOS) imaging technique operating in the near-infrared (NIR) wavelengths, specifically through a raster scanning approach. The results obtained from the experiments demonstrated the effectiveness of this imaging method in recovering images of hidden targets. The imaging setup, which employed a NIR laser with a wavelength of 808 nm and an output power of 500 mW, successfully captured images by reflecting the laser beam off a relay wall and the hidden target before it reached the NIR camera. The experiments conducted with three different targets yielded promising results, showcasing the capability of the proposed method to reconstruct images even when the target was obscured from direct view.The mean squared error (MSE) and root mean square error (RMSE) values calculated during the imaging process indicated that while the accuracy of the recovered images was influenced by the number of scan points, the overall performance remained satisfactory. The trade-off between the number of scan points and the resulting MSE highlighted the importance of optimizing scanning parameters to achieve a balance between image quality and processing time.
\section{AI Usage Statement}
The authors declare that they used artificial intelligence tools only to improve and edit the text of the article.


\end{document}